\documentclass[sigconf,nonacm,screen]{acmart}
\usepackage{amsfonts}
\usepackage{amsmath}
\usepackage{amsthm}
\usepackage{array}
\usepackage{bm}
\usepackage{booktabs}
\usepackage{bxtexlogo}
\usepackage{caption}
\usepackage{enumitem}
\usepackage{float}
\usepackage{geometry}
\usepackage{graphicx}
\graphicspath{{figs/}}
\usepackage[utf8]{inputenc}
\usepackage{makecell}
\usepackage{mathtools}
\usepackage{multirow}
\usepackage{natbib}
\setcitestyle{authoryear,numbers,aysep={}}
\usepackage{soul}
\usepackage{stfloats}
\usepackage{subcaption}
\usepackage[switch]{lineno}
\usepackage{arydshln}
\usepackage{microtype}
\usepackage{balance}
\def\innerprod<#1>{\langle #1 \rangle}

\theoremstyle{definition}

\theoremstyle{remark}

\AtBeginDocument{%
  }

\usepackage{hyperref}

\begin{document}

\title{LOB-ID: Evaluating Synthetic Market Data by Inception Distances}

\author{Andreea Bacalum}
\affiliation{%
  \institution{Simudyne}
  \city{London}
  \country{United Kingdom}}

\author{Zhuohan Wang}
\authornote{Joint affiliation with King's College London, United Kingdom.}
\affiliation{%
  \institution{Simudyne}
  \city{London}
  \country{United Kingdom}}

\author{Ollie Olby}
\affiliation{%
  \institution{Simudyne}
  \city{London}
  \country{United Kingdom}}

\author{Martin Garaj}
\affiliation{%
  \institution{Simudyne}
  \city{London}
  \country{United Kingdom}}

\author{Namid Stillman}
\affiliation{%
  \institution{Simudyne}
  \city{London}
  \country{United Kingdom}}
\email{namid@simudyne.com}

\renewcommand{\shortauthors}{Bacalum et al.}

\begin{abstract}
Generative models of limit orderbook (LOB) data have advanced rapidly, but their evaluation often focuses on stylised facts and selected market statistics. These measures provide useful diagnostics but may not capture the joint temporal and cross-level structure of order-book trajectories. We introduce LOB-ID, an embedding-based framework that adapts the Fr\'echet Inception Distance (FID) and Monge Inception Distance (MIND) to LOB data. To obtain domain-specific embeddings, we train the DeepLOB architecture on four months of Level-2 order-book data for five equities. We show that LOB-ID is stable across time, instruments, and embedding checkpoints, and rises monotonically under controlled distortions. We then construct a moment-matching attack against FID and a deep-book perturbation that evades statistic-based evaluation. MIND remains substantially more sensitive to both distortions. Finally, we score five generative LOB models, spanning stochastic baselines and deep learning approaches, and find that LOB-ID ranks them in line with the joint temporal and cross-level structure each captures by construction.
\end{abstract}




\maketitle

\newcommand{\best}[1]{\bm{#1}}          
\newcommand{\snd}[1]{\underline{#1}}

\section{Introduction}

Synthetic market data is becoming an increasingly useful tool in quantitative finance \cite{assefa2020generating}. It supports applications such as backtesting, execution analysis, risk measurement, and market-microstructure research, and it reduces reliance on historical datasets that can be costly. It also addresses a more basic limitation of historical data: real markets provide only one realised path of what occurs, whereas generative models produce many alternative scenarios, allowing market behaviour to be studied beyond what has already happened.

A limit orderbook (LOB) records outstanding bids and asks at each price level, with a matching engine pairing incoming orders against resting ones by price-time priority \cite{gould2013limit}. The book evolves through a stream of messages, so a synthetic LOB model must generate this order flow rather than prices alone.

Recent generative models have substantially improved the realism of synthetic order flow, but progress in generation has outpaced progress in evaluation \citep{coletta2022towards,nagy2023generative,nagy2025lobbench}. Historical data provides a ground-truth reference, yet evaluation cannot be posed as matching a specific sequence. Models are not expected to perfectly reproduce a single realised market path, but to produce plausible alternatives from the same underlying distribution. Evaluation is therefore a question of distributional similarity rather than point-wise accuracy. Hence, most existing validation methods assess generative models of LOB data by checking whether the generated data reproduce selected properties of real markets. Such comparisons are informative but partial, since a sequence can match a wide range of individual statistics and still fail to behave like a coherent market trajectory.

What these comparisons tend to miss is the underlying joint structure of the data. A realistic LOB sequence should preserve the dependence between successive book states, between liquidity at the touch and in the deeper levels, between the bid and ask sides of the book, and between all of these and the prevailing market regime. Such dependencies are hard to capture with a fixed set of hand-designed statistics, as they are spread across time and price levels, and evaluating them benefits from a representation of the order-book trajectory as a whole.

We address this problem by introducing inception distances for LOB (LOB-ID), an embedding-based framework for evaluating synthetic LOB data. To achieve this, we train a DeepLOB network on four months of Level-2 order-book data for five equities listed on the Hong Kong Stock Exchange (HKEX) \citep{zhang2019deeplob}. After removing the LSTM and classification head, we use the network's Inception representation as a domain-specific feature extractor. We then apply the Fr\'echet Inception Distance (FID) and the Monge Inception Distance (MIND) to the resulting embeddings \citep{heusel2017gans,berthet2026mind}. As the metrics are based on representations constructed from Level-2 order-book windows, they are sensitive to the joint structure that statistic-based comparisons can overlook.

This work makes four contributions. First, we introduce LOB-ID, adapting FID and MIND metrics from image-generation evaluation to representations learned from LOB data. Second, we show that MIND is consistent across nearby trading days, distinguishes between instruments, responds monotonically to controlled perturbations, and stabilises at later stages of DeepLOB training. Third, we analyse the adversarial weaknesses of FID and selected statistic-based LOB evaluations and show that MIND remains sensitive to the loss of joint structure they miss. Fourth, we compare Zero Intelligence, Compound Hawkes, LOBGAN, LOBS5 and DiffLOB on a common dataset using LOB-ID, stylised facts, and existing benchmark metrics. We believe our work will support a more rigorous evaluation of synthetic market
data by assessing the joint structure of the
generated order-book trajectories that standard statistics may miss.
\section{Related Work}
\label{sec:related_work}

\paragraph{Synthetic LOB models}
Early methods for generating synthetic LOB data were typically based on stochastic models. For example, zero-intelligence models generate orders according to exogenously specified stochastic rules, without explicitly modeling strategic optimisation or learning by market participants \citep{farmer2005predictive}. Alternatively, other models use Hawkes processes to allow past events to affect the rate of future events, which have been shown to capture clustered and self-exciting order flow \citep{bacry2015hawkes,jain2024compound}. Although interpretable, these models require the event structure and arrival dynamics to be specified and calibrated in advance.

The development of deep learning shifted the focus towards learning representations directly from LOB data. For example, LOBGAN generates synthetic order flow using a conditional generative adversarial network \citep{coletta2022towards, coletta2023conditional}. LOBS5 instead treats order-flow generation as an end-to-end autoregressive sequence-modelling task \citep{nagy2023generative}. LOBS5 tokenises message-level LOB data and generates the tokens sequentially using simplified structured state-space layers, which can process long histories efficiently. More recently, DiffLOB uses a regime-conditioned diffusion model to generate future LOB trajectories under specified market conditions \citep{wang2026difflob}. It conditions on historical LOB states and future
regimes defined by trend, volatility, liquidity, and order-flow imbalance. This enables controllable and counterfactual generation, allowing the model to
simulate how the orderbook may evolve under hypothetical future conditions.


\paragraph{Evaluation of synthetic LOB data}
Generative LOB models have often been evaluated using so-called `stylised facts' such as heavy-tailed returns, volatility clustering, and persistent order-flow dependence \citep{cont2001empirical,lillo2004longmemory}. These tests show whether a simulator reproduces well-known properties of real markets. However, the choice of statistics and the way they are measured often differ across studies, which makes direct comparison difficult.

More recently, LOB-Bench introduced a standardised framework for evaluating synthetic LOB data \citep{nagy2025lobbench}. It compares real and generated data through distributions of selected market-microstructure statistics, using both the $L_1$ norm and Wasserstein-$1$ distance. The framework also evaluates price-impact responses across multiple lags and includes a trained discriminator that measures how well real and generated trajectories can be separated. Together, these components provide a broad assessment of the realism and quality
of synthetic LOB data. In this work, we use the distributional and price-impact components of LOB-Bench as an evaluation framework and do not directly evaluate its discriminator.

\paragraph{Embedding-based generative metrics}
Embedding-based metrics offer a complementary approach to evaluation. They pass real and generated samples through a fixed, pre-trained network and compare the resulting feature distributions. Similarity is therefore measured in a learned representation space rather than directly from raw inputs or hand-selected summary statistics. Past work has considered embedding networks within the context of calibration for generative models of market data but not the evaluation of the resulting synthetic data \cite{stillman2023deep}. 

The Fr\'echet Inception Distance (FID) was introduced for evaluating generated images by comparing Gaussian approximations of the Inception-v3 feature distributions of real and generated samples \citep{heusel2017gans}. However, FID captures only the first two moments of these distributions, and its empirical estimate can be biased or unstable when the sample size is small \citep{chong2020effectively}.

The Monge Inception Distance (MIND) was introduced as an alternative to FID
for evaluating generative models \citep{berthet2026mind}. Rather than
approximating the embedding distributions as Gaussian, MIND computes a scaled
average of the squared $2$-Wasserstein distances between their
one-dimensional projections along random unit directions. This allows it to
capture differences beyond the first two moments without estimating
high-dimensional covariance matrices. 
\section{Methods}
\label{sec:methods}

In this section, we outline how we adapt the Fr\'echet Inception Distance (FID) and the Monge Inception Distance (MIND) to the evaluation of LOB data. Both metrics compare distributions of feature embeddings produced by the pre-trained network, conventionally the Inception-v3 model trained on ImageNet. Since features learned from natural images are inappropriate for financial data, we replace this backbone with a domain-specific feature extractor trained directly on LOB data. We use the DeepLOB model, whose architecture and training are described in Section~\ref{sec:inception_network}. The trained network yields an embedding function $\psi_w$ that maps a window of $T=100$ successive event-indexed Level-2 book states, $x \in \mathbb{R}^{T \times 40}$, to a fixed $96$-dimensional feature vector. This representation plays the same role as the Inception-v3 features in the image setting, while capturing relationships across price levels and over time that are specific to LOB data.

Both real and generated samples are passed through $\psi_w$ to obtain two sets of embeddings, whose distributions are then compared. FID compares Gaussian approximations of the embedding distributions, whereas MIND compares their random one-dimensional projections. Lower values indicate greater similarity in the learned LOB
representation space.

All experiments use four months of message-level (L3) data collected between September and December 2025 for five liquid equities listed on the Hong Kong Stock Exchange (HKEX): 0700.HK (Tencent), 1024.HK (Kuaishou), 9999.HK (NetEase), 3690.HK (Meituan), and 9888.HK (Baidu). From the raw event streams, we construct the order messages and corresponding Level-2 books containing the ten best price levels on each side. We split the data chronologically, using 1 September to 12 December for training, 13 to 20 December for validation, and 21 to 31 December for testing, so that all evaluation is performed on market days that follow the training period. We choose this period so that evaluation spans a range of liquidity conditions: the training window covers active trading, while the validation and test windows fall in the year-end holiday period, when participation is materially reduced. This provides a natural distribution shift under which to assess how the LOB-ID metric changes in thin markets. The same split and book-reconstruction pipeline are used throughout the experiments, for both the embedding network and the generative models. Although the source dataset contains message-level data and some generators operate at the message level, all inputs to the LOB-ID feature extractor are Level-2 book-state windows. Accordingly, when discussing joint structure in the context of LOB-ID, we refer specifically to temporal, cross-level, and bid--ask dependencies within Level-2 trajectories. LOB-ID does not directly
evaluate the full message-level order-flow process.

\subsection{Synthetic Market Data}
\label{sec:synthetic_data}

We train and evaluate five synthetic LOB models using the common dataset and temporal split described above: Zero-intelligence (ZI), Compound Hawkes process (CH), LOBGAN, LOBS5 and DiffLOB. 

Our implementations of the ZI and CH baselines are adapted from the calibration procedures used in \citet{kawawa2026tradefm}. The ZI model independently samples order side and action type from empirical categorical distributions, inter-arrival times and volumes from fitted exponential distributions, and price depth from a Gaussian mixture model. It therefore preserves calibrated marginals while removing temporal and conditional dependence. The CH model introduces temporal dependence through an eight-dimensional Hawkes process, with one dimension for each action--side combination and sum-of-exponential kernels at four fixed timescales. For each test chunk, it conditions on the preceding $1024$ real events and generates the next $1024$ events using Ogata's thinning algorithm \citep{ogata1981lewis}. Volumes and price depths are sampled from fitted exponential and Gaussian-mixture mark distributions, respectively.

We also consider three deep generative models for LOB data. LOBGAN follows the conditional architecture of \citet{coletta2022towards}. Its generator combines an LSTM representation of the previous $30$ nine-dimensional market states with a $64$-dimensional Gaussian noise vector to generate a seven-feature order representation. During closed-loop training, generated orders update the market-state features fed back to the generator for up to five unrolled steps. LOBS5 is implemented as a masked-token generator over message-level LOB data \citep{nagy2023generative}. Its message and book branches process $22$-token order messages and a $41$-dimensional signed volume image using structured state-space layers. Generation proceeds token by token from a real message context, with each completed message and updated book state fed back into the model. DiffLOB is a regime-conditioned diffusion model that jointly generates future Level-2 price and volume trajectories rather than individual order messages \citep{wang2026difflob}. During training, the conditioning information is randomly dropped with probability $0.5$, allowing the model to learn both conditional and unconditional score estimates. We evaluate DiffLOB in two settings. First, we condition it on the historical LOB and the realised future trend, volatility, liquidity, and order-flow imbalance to test whether it can reproduce a specified market regime. Secondly, to observe how the model performs without conditioning information, we set all conditioning inputs to null, so the model generates unconditionally from the learned distribution.

A central challenge for LOB models is determining whether the resulting synthetic data are realistic. We evaluate each generator using three complementary approaches. First, we test the 11 stylised facts of financial time series considered by \citet{cont2001empirical}. Second, we compute the distributional metrics introduced by LOB-Bench \citep{nagy2025lobbench}, which compare real and generated data using
distances over a standard set of market-microstructure statistics, including
spread, imbalance, timing, and volume measures. The full list of statistics is provided by \citet{nagy2025lobbench}. Since all generators are evaluated on Level-2 book states, we use 12 of the 21 LOB-Bench measures that do not require message-level data. Finally, we introduce LOB-ID, our own method of evaluating synthetic order-book data, an embedding-based evaluation framework using Inception Distances. We provide a comparison of the relative ranking of models with respect to these different evaluation frameworks in Section~\ref{sec:results_synthetic_data}.

\subsection{Inception Distances}

\subsubsection{Inception Network}
\label{sec:inception_network}
An Inception network is a deep convolutional neural network built around the Inception module, introduced in GoogLeNet by Szegedy et al. \citep{szegedy2015going}. The module applies convolutions of multiple sizes in parallel and concatenates their outputs, allowing features to be captured at multiple scales.
\

\indent DeepLOB combines convolutional layers, an Inception module, and an LSTM to predict short-term mid-price movements from raw LOB data~\citep{zhang2019deeplob}. We follow the architecture, preprocessing, and training procedure of \citet{zhang2019deeplob}, using the dataset and temporal split described above. Each input contains the most recent $T=100$ order-book events, $x \in \mathbb{R}^{T \times 40}$, with price and size information from the ten best bid and ask levels, producing a 96-dimensional feature vector.


An LSTM and linear classification head use these features to predict whether the mid-price will move down, remain stationary, or move up over the next $100$ messages. On the held-out test set, the trained DeepLOB classifier achieves an accuracy of $75.44\%$ and a macro-$F_1$ score of $75.29\%$, indicating balanced predictive performance across the three mid-price movement classes. For LOB-ID, we remove the recurrent and classification layers and use the time-averaged output of the Inception module as the embedding, namely

\begin{equation}
\psi_w(x) = \frac{1}{T} \sum_{t=1}^{T}
\operatorname{Inception}
\!\bigl(\operatorname{Conv}(x)\bigr)_{:,t}
\in \mathbb{R}^{96}.
\end{equation}

\subsubsection{Fr\'echet Inception Distance (FID)}

The Fr\'echet Inception Distance (FID) compares real and generated embeddings
in the feature space of a fixed network \citep{heusel2017gans}. Let
$p_{\mathrm{data}}$ and $p_{\theta}$ denote their respective distributions.
FID fits Gaussians with empirical means $\mu,\tilde{\mu}$ and covariance
matrices $\Sigma,\tilde{\Sigma}$, and computes
\begin{equation}
\mathrm{FID}(p_\theta,p_{\mathrm{data}})
=
\lVert \mu-\tilde{\mu} \rVert_2^2
+
\operatorname{tr}\!\left[
\Sigma+\tilde{\Sigma}
-2\left(
\Sigma^{1/2}\tilde{\Sigma}\Sigma^{1/2}
\right)^{1/2}
\right].
\label{eq:fid}
\end{equation}
A lower FID indicates greater similarity between the real and generated embedding distributions. However, the Gaussian approximation means that FID depends only on their means and covariances and therefore cannot detect differences beyond the first two moments.

\subsubsection{Monge Inception Distance (MIND)}
\label{subsec:mind}

The Monge Inception Distance (MIND) is an embedding-based measure introduced
as an alternative to FID \citep{berthet2026mind}. Rather than fitting Gaussian
approximations to the embedding distributions, MIND averages the squared
$2$-Wasserstein distances between their one-dimensional projections along
random unit directions. For embeddings in $\mathbb{R}^{96}$ and a unit
direction $u$ drawn uniformly from the sphere $\mathbb{S}^{95}$, MIND is
defined as
\begin{equation}
\mathrm{MIND}(p_\theta,p_{\mathrm{data}})
=
\alpha\,
\mathbb{E}_{u\sim\mathcal{U}(\mathbb{S}^{95})}
\!\left[
W_2^2
\left(
u^\top p_\theta,
u^\top p_{\mathrm{data}}
\right)
\right].
\label{eq:mind}
\end{equation}

Here, $u^\top p$ denotes the distribution of $u^\top X$ for $X\sim p$, and
$W_2^2$ is the squared $2$-Wasserstein distance. Following the empirical scaling proposed by \citet{berthet2026mind}, we use $\alpha=3d=288$, where
$d=96$ is the embedding dimension. Note that this construction differs from
the Wasserstein-$1$ distance used by LOB-Bench \citep{nagy2025lobbench},
which compares univariate distributions of individual scalar market
statistics rather than projections of the embedding distribution. All
reported MIND values estimate the expectation in Eq.~\eqref{eq:mind} using
$256$ random projections.

For uniformly weighted empirical distributions of equal size, each
one-dimensional squared $2$-Wasserstein distance can be computed exactly by
sorting the projected values. MIND therefore avoids estimating and storing
high-dimensional covariance matrices. At the population level, MIND is zero
if and only if the two embedding distributions are equal. Matching only their
means and covariances is therefore insufficient to make MIND zero.
\citet{berthet2026mind} show empirically that MIND is more resistant, though
not immune, to moment-matching attacks than FID. We confirm this behaviour for
LOB embeddings in Section~\ref{sec:moment_matching}, where MIND remains
substantially more sensitive than FID after the embedding moments are
approximately matched. We therefore use MIND as the primary score within
the LOB-ID framework throughout the remainder of the paper.

\section{Results}
\label{sec:results}

In this section, we show that LOB-ID behaves as a reliable evaluation metric for synthetic market data: it is consistent across time and instruments, responds monotonically to perturbations, stabilises across embedding checkpoints, and remains sensitive to adversarial attacks that FID and selected LOB-Bench statistics substantially underestimate or miss. We then use it to score the generative models of Section~\ref{sec:synthetic_data}. All experiments use the trained DeepLOB Inception layer as the embedding network. The stability analysis and the moment-matching attack use all five
stocks over the six test days, since they require only real data. Experiments involving the generative models, namely the joint-structure attack, the checkpoint analysis, and the model comparison, use three stocks (9999.HK, 1024.HK, 0700.HK), as each generator must be separately trained and calibrated per instrument.


\subsection{Stability and Robustness of LOB-ID}

\begin{figure}[!t]
    \centering
    \includegraphics[width=0.9\linewidth]{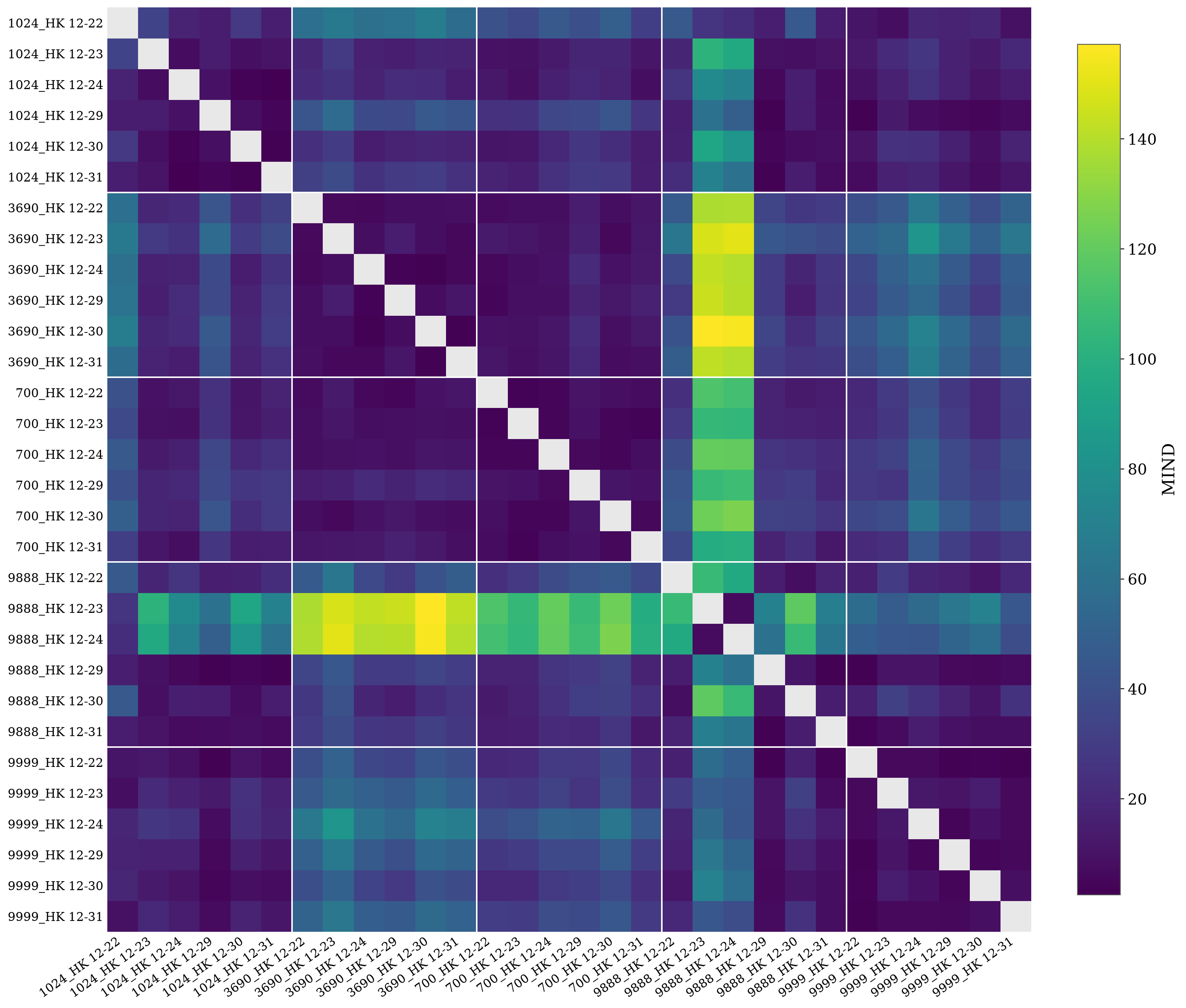}
    \caption{Pairwise MIND distances between all stock-days in the test period
(five stocks, six dates each). Each $6\times6$ diagonal block compares
different days of the same stock, while the off-diagonal blocks compare
different stocks. Darker colours indicate smaller distances; the white main
diagonal marks each stock-day compared with itself, where MIND is zero.} 
    \label{fig:heatmap}
\end{figure}

We expect samples from the same stock taken on nearby days to have similar LOB structure. We therefore compare MIND across all stock--day pairs in the test period. Figure~\ref{fig:heatmap} shows the pairwise MIND distances. The white entries on the main diagonal are zero because each stock--day is compared with itself. The off-diagonal entries within each $6 \times 6$ stock block are consistently dark, showing that MIND assigns low distances to different days from the same
instrument. In contrast, the cross-stock blocks are substantially lighter, revealing a clear separation between instruments. A small number of dates depart from this pattern and show larger distances from nearby days, particularly around the Christmas period.

\subsection{Sensitivity Analysis}
\label{sec:perturbations}

We test whether MIND responds to small, gradual corruptions by applying perturbations of increasing strength $\varepsilon$ to the historical LOB data. Each input window is $X \in \mathbb{R}^{T \times 40}$, where $T=100$ and the features contain the $z$-scored prices and volumes at the ten best bid and ask levels. We modify the price path directly as well as the temporal structure of the window (through deletions).

Let $\mathcal{P}$ denote the set of $20$ price columns. For each price perturbation, we apply the same time-dependent shift $\delta_t$ to every price
level,
\begin{equation}
\widetilde{X}_{t,j}
=
X_{t,j}+\sigma_p\delta_t,
\qquad
j\in\mathcal{P},
\label{eq:price-perturbation}
\end{equation}
where $\sigma_p$ is the mean standard deviation of the price features. The volume columns are unchanged. We draw $s\sim\operatorname{Unif}\{-1,+1\}$ independently for each window so that the perturbations are approximately zero-mean across the dataset.

For $t\in\{0,\ldots,T-1\}$, the three price shifts are
\begin{align}
\delta_t^{\mathrm{trend}}
&=
\varepsilon s
\left(
\frac{2t}{T-1}-1
\right),
\label{eq:trend-perturbation}
\\
\delta_t^{\mathrm{walk}}
&=
\varepsilon\widehat{W}_t,
\label{eq:walk-perturbation}
\\
\delta_t^{\mathrm{jump}}
&=
\varepsilon s\,
\mathbf{1}\{t\geq\tau\},
\qquad
\tau\sim\operatorname{Unif}[T/4,3T/4).
\label{eq:jump-perturbation}
\end{align}
The linear trend creates a total upward or downward movement of
$2\varepsilon\sigma_p$ across the window. For the random-walk perturbation, $\widehat{W}$ is a Gaussian
random walk standardised to zero mean and unit variance within each window.
It introduces smooth stochastic movement with scale
$\varepsilon\sigma_p$. The jump perturbation introduces a single step change
at a randomly selected point in the middle half of the window.

The fourth perturbation, row deletion, removes each row independently with probability $\varepsilon$. The surviving rows retain their order and are shifted to the beginning of the window, while the remaining tail is filled by repeating the final surviving row. Successive retained rows are therefore separated by $1/(1-\varepsilon)$ original events on average. Thus, $\varepsilon$ acts on the event index rather than on feature values, and we evaluate this perturbation over a smaller range.


\begin{figure*}[!t]
    \centering
    \includegraphics[width=\linewidth]{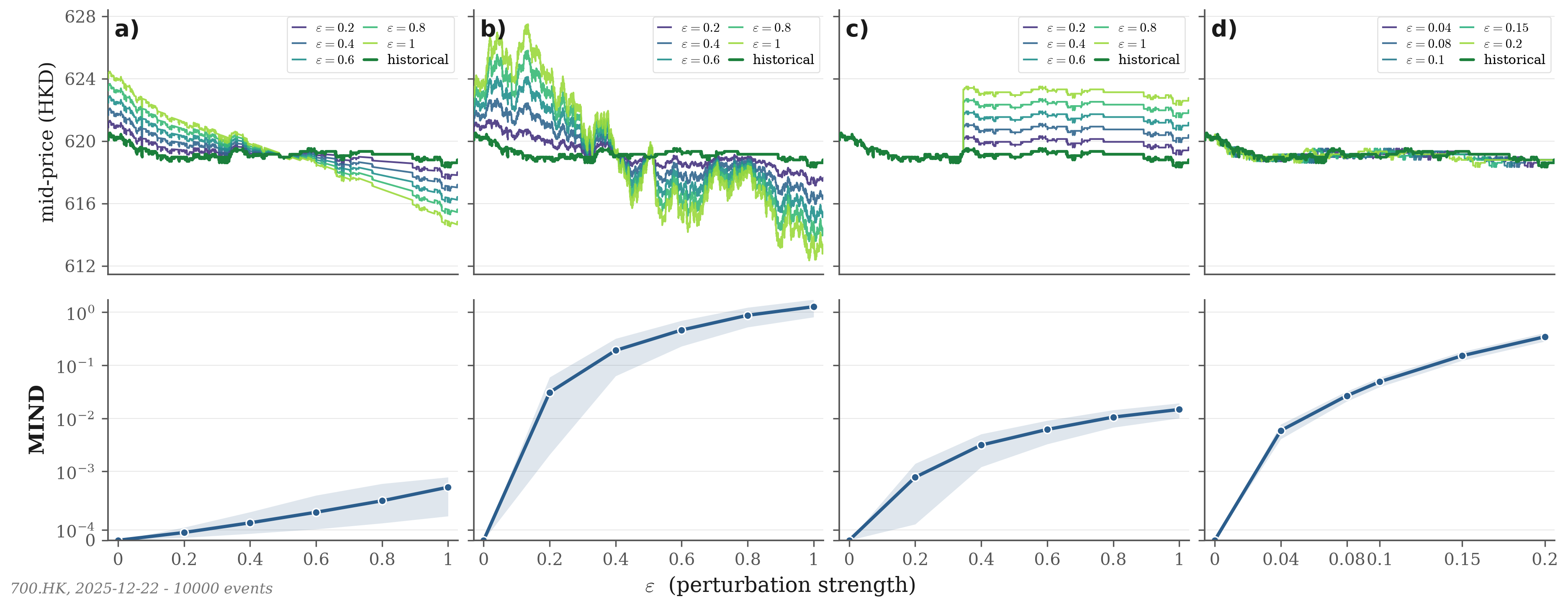}
    \caption{MIND as a function of perturbation strength $\varepsilon$ for the
four perturbation mechanisms: a)~linear trend, b)~random walk, c)~price
jump, and d)~row deletion. Top: example mid-price paths for a single
stock-day (0700.HK, 22 December 2025) at increasing $\varepsilon$, with the
historical unperturbed path in dark green. Bottom: mean MIND (log scale) across
stock-days, with shaded bands showing $\pm 1$ standard deviation.}
    \label{fig:price-perturb}
\end{figure*}

Figure~\ref{fig:price-perturb} shows the mean MIND score for the four perturbation mechanisms, with shaded bands representing one standard deviation across stock-days. For all four perturbations, MIND is zero when $\varepsilon=0$ and rises steadily as the distortion becomes stronger, with no reversals. The perturbations are deliberately kept subtle to test whether MIND can detect small deviations from the real data. Accordingly, the MIND score increases only moderately for the trend and jump perturbations. The random-walk perturbation introduces more persistent noise throughout each price level and therefore produces a larger increase in MIND. Row deletion has a similarly strong effect because removing events disrupts the temporal structure and spacing of the original order-book sequence.

\subsection{Adversarial Attacks}

\subsubsection{Moment-Matching Attack on FID}
\label{sec:moment_matching}

FID compares embedding distributions using only their means and covariance matrices \citep{heusel2017gans}. It therefore cannot distinguish distributions that share these moments but differ in shape. Motivated by \citet{berthet2026mind}, we construct an attack in the LOB input space that maximises the difference from the real embedding distribution while keeping its mean and covariance close to the real values.

We collect up to $20{,}000$ real embeddings from the five stocks over the test period and, after shuffling with a fixed random seed, divide them equally into a target pool $X_A$ and an evaluation pool $X_B$. The target mean $\mu$ and covariance $\Sigma$ are estimated from the full pool $X_A$.

The attack starts from $n=1024$ real LOB windows and treats the input tensor
$a \in \mathbb{R}^{n\times T\times40}$ as the optimisation variable while
keeping the DeepLOB network fixed. Writing $E=\operatorname{embed}(a)$, we
maximise the shape loss $\mathcal{L}_{\mathrm{shape}}(a)$ subject to
$\mathcal{L}_{\mathrm{mom}}(a)\leq\tau$. The shape loss averages the squared
$2$-Wasserstein distances between one-dimensional projections of $E$ and a
fixed $n$-sample subset of $X_B$, using $K=128$ random directions redrawn at
each optimisation step. The moment loss is
\begin{equation}
\mathcal{L}_{\mathrm{mom}}(a)
=
\left\|\bar{E}-\mu\right\|_2^2
+
\frac{1}{d}
\left\|\Sigma_E-\Sigma\right\|_F^2,
\qquad d=96,
\end{equation}
where $\bar{E}$ and $\Sigma_E$ are the mean and covariance of the attacked
embeddings, and $\mu$ and $\Sigma$ are the corresponding target moments
estimated from $X_A$. The threshold $\tau$ is the moment discrepancy of a
genuine real batch of size $n$, so a feasible attack matches the real moments
within the sampling variation of a genuine batch.

We optimise in alternating distortion and polishing phases. The distortion phase increases the shape loss using Adam with learning rate $5\times10^{-2}$, while an adaptive penalty keeps the moment loss below $2\tau$. The polishing phase minimises only the moment loss, starting with a learning rate of $2.5\times10^{-2}$ and reducing it on plateaus to $10^{-4}$. Polishing stops when the loss falls below $0.2\tau$. We run three cycles and retain the feasible result with the largest MIND score.

Since empirical FID and MIND depend on sample size, all final comparisons use matched batches of $n=1024$ embeddings. Let $X_A^{(n)}\subset X_A$ and $X_B^{(n)}\subset X_B$ denote the real batches used to compute the real--real noise floor. For each metric $m$, we report
\begin{equation}
R_m =
\frac{
d_m\!\left(E,X_B^{(n)}\right)
}{
d_m\!\left(X_A^{(n)},X_B^{(n)}\right)
}.
\label{eq:attack-ratio}
\end{equation}
Here, $E$, $X_A^{(n)}$, and $X_B^{(n)}$ each contain $1024$ embeddings. A ratio close to one means that the attacked samples appear as similar to the real data as two genuine real batches, while a larger value indicates that the attack remains detectable.

The optimisation converges with a moment loss of $0.0033$, satisfying $\mathcal{L}_{\mathrm{mom}}\leq\tau$. FID increases from a real--real noise floor of $0.17$ to $0.59$, giving $R_{\mathrm{FID}}=3.5$. MIND increases from $0.54$ to $40.3$, yielding $R_{\mathrm{MIND}}=74.5$, approximately $21$ times the FID ratio. Thus, FID does not miss the attack entirely but substantially understates the remaining distributional distortion, whereas MIND places the attacked data far outside the real--real baseline.




\begin{figure*}[!t]
    \centering
    \includegraphics[width=0.95\linewidth]{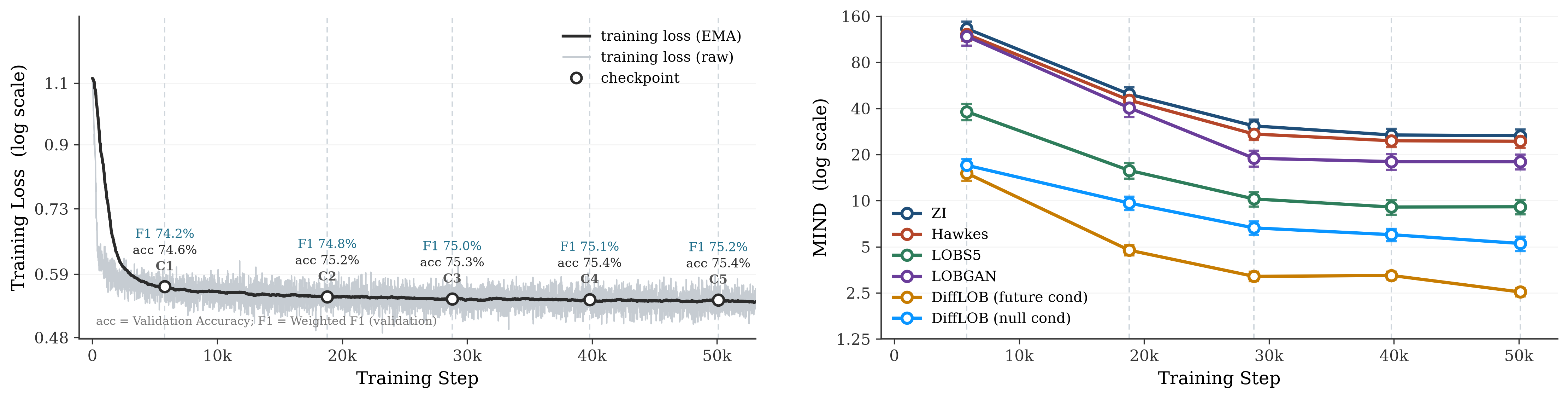}
    \caption{Effect of inception-network training on LOB-ID. \textbf{Left:}
DeepLOB training loss (log scale) over training steps, with five
evaluation checkpoints C1--C5 at different stages of training. \textbf{Right:} MIND (log scale) between historical data
and synthetic data from each one of five generative models, computed at the same five
checkpoints; error bars show $\pm 1$ standard error of the mean.} 
    \label{fig:loss_vs_mind}
\end{figure*}

\subsubsection{Joint-Structure Attack against Statistic-Based LOB Evaluation}

LOB-Bench evaluates synthetic LOB data using distributions of selected market statistics, price-impact responses computed from the mid-price, and a trained discriminator \citep{nagy2025lobbench}. Here, we focus on the first two components. Marginal distributional comparisons do not verify that deep-book liquidity evolves coherently over time, with the Level-$1$ state, with the opposite side of the book, or with the prevailing market regime. We therefore construct adversarial LOB data that preserve the evaluated marginals and impact response while deliberately breaking these dependencies. We omit the discriminator because it serves a different evaluation objective: measuring the empirical separability of real and generated sequences. This experiment instead isolates whether the evaluated statistics and impact response detect the perturbation. We leave discriminator-based evaluation of the attack for future work. 


We construct the attack for each real trading day. Prices, timestamps, message features, and quantities at level $1$ are left unchanged, while the size and order count columns at levels $2$ to $10$ are rearranged. For each side of the book, we divide the data into contiguous segments separated by price changes in the deep levels and permute only segments of equal length. Reordering the deep-book values does not change their marginal distributions. Most LOB\nobreakdash-Bench statistics use only Level-$1$ data, which the attack leaves unchanged, so their scores are unaffected. For every stock-day we report each metric as its percentage above that day's real--real noise floor, obtained by splitting the day into even and odd row blocks: $-100\%$ means the metric sees no difference at all, $0\%$ is the sampling floor, and positive values mean detection.


We find that the distributional statistics and impact-response components from LOB-Bench do not detect it. Averaged over its $12$ distributional statistics that do not require message-level data, the attack sits $95\% \pm 1\%$ below the noise floor across the $18$ stock-days. This follows from the construction. Indeed, $9$ of the $12$ statistics are preserved exactly, whereas the three that shift (total bid volume, total ask volume, and volume per minute) stay below their real--real
baselines, and the impact response function is untouched. Every evaluated LOB-Bench statistic rates the adversarial sequence at least as favourably as a second genuine sample of the same day. However, we find that LOB-ID, using MIND, is able to detect the attack. Averaged over the same $18$ stock-days, the attack sits $75\% \pm 17\%$ above the noise floor, and the mean is positive for every
stock: $+40\%$ for 9999.HK, $+144\%$ for 1024.HK, and $+42\%$ for 0700.HK. LOB-ID consistently places the attack above genuine sampling variation.

\subsection{Impact of Inception Network Training}
We next examine how the training stage of the DeepLOB Inception network affects the stability and consistency of the MIND score when applied to different LOB generative models. We select five checkpoints at progressively later stages of training and, for each, compute LOB-ID between the historical data and synthetic data generated by five fixed models (ZI, CH, LOBGAN, LOBS5 and DiffLOB with null and future conditioning). Since the generators are held constant throughout, any change in LOB-ID across checkpoints reflects the evolving embedding network rather than a change in the generated data.

As shown in Figure~\ref{fig:loss_vs_mind}, as DeepLOB training approaches convergence, the MIND scores also converge towards stable plateaus across all generator configurations. Most of the reduction occurs at the earlier checkpoints, while the changes between approximately $40{,}000$ and $50{,}000$ training steps are small. This indicates that, once the DeepLOB representation is sufficiently trained, the resulting MIND scores become less sensitive to the particular checkpoint used to construct the embedding space.

\subsection{Comparison of Generative Models}
\label{sec:results_synthetic_data}

We compare five generative models using stylised facts, LOB-Bench
distances, and LOB-ID, evaluating DiffLOB under both null and future
conditioning. Table~\ref{tab:model-comparison} reports the mean $\pm$ one
standard deviation across three stocks and six test days. The variation
therefore reflects differences across stocks and dates. We separate future-conditioned DiffLOB from the other generators in Table~\ref{tab:model-comparison} because it is an oracle: the trend, volatility, liquidity, and order-flow imbalance variables it conditions on are computed from the realised future trajectory, so it is scored on a strictly easier task. We therefore report it as an upper reference point rather than as a competitor, and rank the remaining models among themselves. Future-conditioned DiffLOB attains the best values on all three distance measures, with low variability across stocks and dates, and reproduces $7.22 \pm 1.00$ stylised facts. Historical data scores $7.67 \pm 0.59$ under the same evaluation.

Among the five non-oracle generators, null-conditioned DiffLOB is strongest
overall. It reproduces the most stylised facts, with $7.33 \pm 0.69$, and
records the lowest $L_1$ and LOB-ID distances, ranking second only on
Wasserstein-$1$. LOBS5 follows, with the lowest Wasserstein-$1$ distance and
second place under both $L_1$ and LOB-ID. The separation is clearest under
LOB-ID, which orders the models as DiffLOB (null), LOBS5, LOBGAN, CH and ZI. Among LOBGAN, CH and ZI, LOBGAN performs best under both LOB-ID and Wasserstein-$1$. The metrics nevertheless produce different rankings, where CH reproduces $6.06 \pm 0.87$ stylised facts, more than the other non-DiffLOB models, but records the highest
Wasserstein-$1$ distance, while LOBS5 reproduces fewer stylised facts while
performing strongly on all three distance measures.

\begin{table*}[t]
    \centering
\caption{Comparison of synthetic LOB generators. Values are means $\pm$ one
standard deviation across $18$ stock-days; lower is better except for stylised
facts. Future-conditioned DiffLOB$^{\dagger}$ uses realised future regime
variables and is excluded from the non-oracle ranking. Best and second-best
non-oracle results are shown in \textbf{bold} and 
\underline{underlined}.}
    \label{tab:model-comparison}
    \renewcommand{\arraystretch}{1.0}
    \begin{tabular}{l c c c c}
        \toprule
        Model & Stylised Facts (of 11) & LOB-Bench (Wasserstein-$1$) & LOB-Bench ($L_1$) & LOB-ID (MIND) \\
        \midrule
        Zero Intelligence & $4.56 \pm 1.82$ & $1.5315 \pm 0.3679$ & $0.6158 \pm 0.0235$ & $26.5514 \pm 11.3785$ \\
        Compound Hawkes   & $\snd{6.06 \pm 0.87}$ & $2.0498 \pm 0.5278$ & $0.5170 \pm 0.0433$ & $24.4254 \pm 9.8464$ \\
        LOBGAN            & $5.67 \pm 1.28$ & $1.1038 \pm 0.2484$ & $0.5642 \pm 0.0903$ & $17.9711 \pm 8.6250$ \\
        LOBS5             & $5.28 \pm 1.78$ & $\best{0.7515 \pm 0.4162}$ & $\snd{0.4195 \pm 0.0450}$ & $\snd{9.1201 \pm 4.2673}$ \\
        DiffLOB (null cond) & $\best{7.33 \pm 0.69}$ & $\snd{0.9168 \pm 0.7167}$ & $\best{0.2735 \pm 0.0776}$ & $\best{6.0251 \pm 3.7417}$ \\
        \hdashline[2pt/2pt]
        DiffLOB (future cond)$^{\dagger}$ & $7.22 \pm 1.00$ & $0.2124 \pm 0.0683$ & $0.1558 \pm 0.0238$ & $2.8204 \pm 0.9424$ \\
        \bottomrule
    \end{tabular}
\end{table*}

\section{Discussion}
\label{sec:discussion}

Our results show that LOB-ID provides a stable measure of similarity between real and synthetic
LOB data. The pairwise comparisons show that the trained embeddings capture persistent
differences between instruments. Samples from the same stock remain close
across nearby dates, while different stocks are clearly separated. The unusually high distances for a small number of dates may reflect changes in liquidity, volatility, or trading activity. We do not investigate the cause here, but the result suggests that LOB-ID could also be used to study changes in market conditions, unusual trading days, or possible regime shifts.

The perturbation experiments support the same conclusion. The changes are deliberately small and remain within the range of normal market variation, yet LOB-ID responds before the altered price paths become visibly unrealistic. Its consistent increase across several different perturbations shows that the
metric is sensitive to a broad range of structural changes, not only to one specific distortion or to changes that are obvious in the raw data.

The checkpoint experiment indicates that, although the absolute MIND scores depend on the stage of DeepLOB training, the relative ordering of the generator configurations remains unchanged across checkpoints. Once the embedding network is sufficiently trained, the scores stabilise and become less sensitive to checkpoint selection. 

The two adversarial experiments expose different weaknesses in FID and the evaluated statistic-based components of LOB-Bench. The first moment-matching attack targets FID directly. Because FID depends only on the embedding mean and covariance, it substantially understates the distortion once these moments are matched. MIND still detects a large difference because it compares the projected
distributions rather than summarising the embeddings only by their means and
covariances.

The joint-structure attack shows a similar limitation in statistic-based benchmarks. The attacked sequence preserves most of the measured marginal distributions and leaves the market-impact response unchanged, but it no longer forms a coherent market trajectory. The attack breaks consistency between the deep-book trajectory and the corresponding order flow, market conditions, and Level-1 state. Although LOB-ID does not observe the message stream directly, it detects the resulting distortion in the temporal, cross-level, and bid–ask structure of the Level-2 book trajectory. Independently rearranging the bid and ask sides also breaks their dependence, while the segment boundaries disrupt continuity in the deep book.

The attacked data therefore appear realistic when individual features are examined separately, but not when the orderbook is viewed as a joint time
series. This supports using the two approaches together. Statistic-based benchmarks provide clear and interpretable diagnostics, while LOB-ID captures temporal and cross-level relationships that individual statistics may miss. Beyond complementing statistic-based diagnostics, LOB-ID provides a reusable distance-based alternative to the LOB-Bench discriminator. The discriminator assigns each trajectory a probability of being real, with model quality inferred from how well its outputs separate real and generated samples. In contrast, LOB-ID directly computes a non-negative discrepancy between the real and generated embedding distributions, without requiring a classification threshold. Once the DeepLOB feature extractor is fixed, the same representation and scoring rule can be applied across generators without training an additional real-versus-generated classifier for each comparison.

Finally, LOB-ID produces an ordering consistent with the dependencies represented by
the tested models. These mechanisms are not observed directly, but they leave signatures in the temporal and cross-level structure of the book trajectories, which is what LOB-ID measures. Zero Intelligence samples side, type, timing, volume, and depth independently, so no relationship between order flow and book state can arise, and it scores worst. Compound Hawkes lets past events raise the intensity of future arrivals, which introduces dependence in event timing, but volumes and depths remain independently sampled, so the
book state is still not explained by its order flow. LOBGAN improves on both because its generator conditions each new order on the current market state, creating genuine flow to book dependence, but that state enters only as a small set of hand-crafted order and book features, so the dependence it can learn is limited to what this summary exposes. LOBS5 improves further because it models the raw message stream directly. Its S5 backbone, a structured state space model, maintains a compressed continuous state of a long message context. DiffLOB models the full trajectory jointly across time and price levels, allowing cross-level and bid--ask dependencies to be represented directly rather than carried implicitly through sequential generation. When the other metrics produce a different ordering, the disagreement reflects differences in the properties emphasised by each
metric. For CH, its large Wasserstein distance may reflect the more extreme timing distributions produced by self-excitation. Future-conditioned DiffLOB receives regime variables computed from the future trajectory, giving it information unavailable to the other generators considered. Its low distances therefore measure regime reproduction rather than unconditional generative quality. However, null-conditioned DiffLOB receives no future information. LOB-ID clearly distinguishes these settings, demonstrating that LOB-ID captures both the benefit of regime information and the underlying quality of the generator.


LOB-ID depends on the choice of embedding network. Its absolute scores are
therefore comparable only when the same representation, preprocessing, sample
size, and projection protocol are used. As the current embedding is constructed from Level-2 book states, it evaluates temporal and cross-level trajectory structure rather than the full message-level process. Our experiments are limited to five liquid HKEX equities, so validation across markets, instruments, and regimes remains future work, although the stable ranking across checkpoints suggests limited sensitivity to checkpoint selection within the chosen representation. A direct empirical comparison between inception distances and trained
discriminator scores remains an important direction for future work.

\section{Conclusion}

We introduced LOB-ID, an embedding-based framework for evaluating synthetic limit orderbook data. By comparing real and generated limit order-book data in the representation space learned by the DeepLOB Inception network, LOB-ID provides a summary of temporal and cross-level similarity between Level-2 order-book trajectories. Our experiments demonstrate that LOB-ID is consistent across instruments and trading days, detects subtle perturbations of the data, and yields stable model rankings. Moment matching substantially reduces FID's sensitivity, while a deep-book attack preserves the selected LOB-Bench statistics. MIND remains sensitive to both distortions. These results highlight LOB-ID as an effective method for measuring joint temporal and cross-level structure in Level-2 trajectories.

\bibliographystyle{ACM-Reference-Format}
\bibliography{bib} 


\end{document}